\documentclass[a4paper,twocolumn,prb]{revtex4}

\usepackage[english]{babel}
\usepackage{latexsym}
\usepackage{graphicx}
\usepackage{amsmath,amssymb}

\begin{document}
\newcommand{\uc}{{\mathrm{c}}}
\newcommand{\uD}{{\mathrm{D}}}
\newcommand{\cs}{C$_{60}$}
\newcommand{\is}{{\em in situ}}
\newcommand{\xs}{{\em ex situ}}

\title{Direct observation by resonant tunneling of the B$^+$
level in a delta-doped silicon barrier}

\author{J.~Caro}
\author{I.D.~Vink}
\author{G.D.J.~Smit}
\author{S.~Rogge}
\author{T.M.~Klapwijk}

\affiliation{Department of NanoScience, Delft University of
Technology, Lorentzweg~1, 2628~CJ Delft, The Netherlands}

\author{R.~Loo}
\author{M.~Caymax}

\affiliation{IMEC, Kapeldreef 75, B-3001 Leuven, Belgium}

\begin{abstract}
We observe a resonance in the conductance of silicon tunneling
devices with a $\delta-$doped barrier. The position of the
resonance indicates that it arises from tunneling through the
B$^+$ state of the boron atoms of the $\delta-$layer. Since the
emitter Fermi level in our devices is a field-independent
reference energy, we are able to directly observe the diamagnetic
shift of the B$^+$ level. This is contrary to the situation in
magneto-optical spectroscopy, where the shift is absorbed in the
measured ionization energy.
\end{abstract}

\date{\today}
\maketitle

The smallest semiconductor device with potential functionality is
a semiconducting nanostructure with a single dopant atom. The
properties of such a structure are most prominent at low
temperature, where the electron or hole is localized at the parent
donor or acceptor, respectively. Manipulation of the wave function
of the charge carrier at the dopant atom with the electric field
of a gate is the obvious tool to influence the properties of the
nanostructure. A good first order description of the properties of
dopant atoms is given by the hydrogen model~\cite{Ramdas}. The
resulting Bohr radius of up to about 10 nm sets the size of a
dopant atom and thus of a single-dopant-atom functional device.
This radius is much larger than for the hydrogen atom, due to
scaling for the effective mass of the carrier and the dielectic
constant of the semiconductor. Such dimensions are accessible with
e-beam lithography and scanning probe techniques. This opens the
way for what can be called atomic scale electronics inside a
semiconductor. Silicon is very attractive for this purpose because
of its highly developed fabrication technology.

A beautiful example of atomic scale electronics inside silicon is
the quantum computer proposed by Kane~\cite{Kane}~and by Skinner
\emph{et al.}~\cite{Kane1}. Both the fabrication and electrical
operation of the qubits of this computer rely on control at the
level of individual phosphorus donors. The highly controlled
dopant engineering required for atomic scale electronics inside
silicon is being worked on by several
groups~\cite{Tucker,Schenkel}. Doping at the atomic scale for
application purposes is thought to be feasible, but development of
this technique will be time consuming.

A more direct way to a single dopant atom in silicon, albeit less
controlled concerning exact positioning, is to use $\delta-$doping
and conventional nanostructuring. Single-dopant-atom structures
fabricated in this way will yield physics relevant for future
devices fabricated with atomic scale doping techniques. To some
extent this approach has been followed already, by including a
$\delta-$layer of dopant atoms in the well of double barrier
diodes~\cite{Lok}. Here, we report a transport study on
$\delta-$doped silicon tunneling devices grown with a single
barrier. The dopant atoms induce zero-dimensional atomic quantum
wells, giving many identical double barrier systems in parallel.
This device, a precursor of a single-dopant-atom device of a
geometry close to a qubit of Kane's computer, shows very
interesting transport properties. In particular, we find a
conductance resonance due to tunneling through the boron
impurities of the $\delta-$layer in the barrier. The position of
the resonance and its magnetic field dependence indicate that it
originates from tunneling through the B$^+$ state of the boron
impurities. So far, this state has only been observed in
spectroscopic studies using photons or phonons and not in an
energy resolved transport experiment like ours.

We fabricate the $\delta-$doped devices from a layered structure
of the type p$^+$ Si(500 nm)/p$^-$ Si(20 nm)/$\delta$/p$^-$ Si(20
nm)/p$^+$ Si(500 nm). Boron is the dopant, for the layers and the
$\delta-$spike. The structure is deposited by chemical vapor
deposition in an ASM Epsilon 2000 reactor on a Si(001) substrate
with low doping, using SiH$_4$ and B$_2$H$_6$ as precursor gasses.
The $\delta-$spike of areal density $1.7\times10^{11}$~cm$^{-2}$
is centered in the lowly doped 40 nm thick tunnel barrier. The
p$^+$ layers are degenerately doped ($N_B=10^{19}$~cm$^{-3}$) and
serve as contact layers. The devices are square mesas, 100, 200,
300 and 400 $\mu$m at a side. They are dry-etched in an SF$_6$
plasma. The etch mask is the Al-1\%Si top contact of the mesa,
which is sputter-deposited through a shadow mask. The SF$_6$ etch
is stopped just after the bottom p$^+$ layer has been reached. A
second shadow mask, aligned with respect to the mesas, is used for
sputter deposition of Al-1\%Si contacts to the bottom layer. The
final step is a 400~$^\circ$C anneal in N$_2$/H$_2$ of the Al
contacts to the p$^+$ Si, using rapid thermal processing. A high
device quality is apparent from resistance scaling with mesa size.
At room temperature the resistance is dominated by the
two-dimensional spreading resistance of the bottom layer between
the mesa and the Al contact, while at low temperature it is
determined by the barrier in the mesa. These devices, which have a
metal-insulator-metal structure, are the simplest all-silicon
tunneling devices.

\begin{figure}[t]
\includegraphics[width=6.5cm]{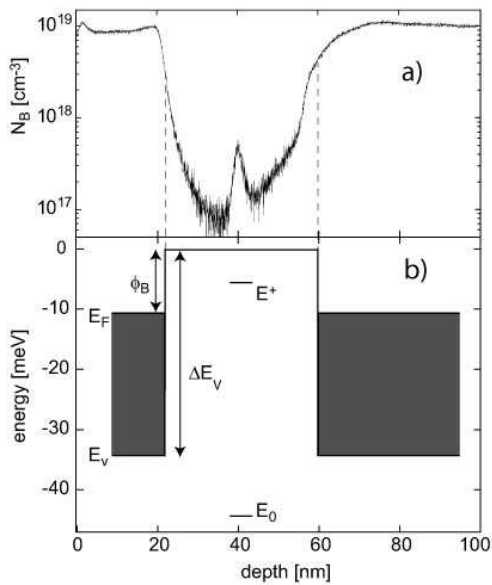}
\caption{(a) SIMS profile of the boron concentration. Zero depth
is the surface position. For optimum resolution the top layer is
only 20 nm thick. Contact layers, barrier and $\delta-$layer are
clearly visible. (b) Depicts the profile of the valence-band edge,
the hole energy increasing in the upward direction. Fermi seas of
emitter and collector extend up to the barrier, of which the
thickness is defined with the criterion
$N_B=4\times10^{18}$~cm$^{-3}$, the concentration of the
metal-insulator transition. Symbols are discussed in the text.}
\label{fig:SIMS}
\end{figure}

We measured the doping profile in the structure with secondary ion
mass spectroscopy (SIMS). Figure~\ref{fig:SIMS}(a) gives the
result, in which the $\delta-$layer, the barrier and the contact
layers are clearly discernible. The $\delta-$layer is about 6 nm
wide and has a peak concentration of
$N_B=5\times10^{17}$~cm$^{-3}$. The contrast of the $\delta-$layer
and the background doping ($N_B\approx 10^{17}$~cm$^{-3}$) in the
barrier is somewhat weak, although for the barrier B$_2$H$_6$ is
only applied during $\delta-$doping. This is due to boron
diffusion during the silicon growth, out of the p$^+$ bottom layer
and out of the $\delta-$layer. The profile of the valence band
edge is depicted in Fig.~\ref{fig:SIMS}(b). The tunnel-barrier
height is~\cite{Yuan} $\phi_B=\Delta E_v-E_F$,~\emph{i.e.} the
valence-band contribution $\Delta E_v$ to the band-gap narrowing
of emitter and collector minus the Fermi energy $E_F$ of these
device layers.

\begin{figure}[t]
\includegraphics[width=6.5cm]{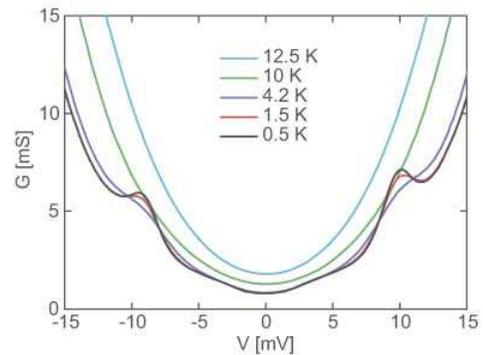}
\caption{Conductance curves of 400 $\mu$m device 1/888/1, for the
temperatures listed. With decreasing temperature the conductance
decreases (for the lower temperatures most clear at higher biases)
and a resonance at $\pm$10 mV develops. Between 1.5 and 0.5 K the
resonance still grows.} \label{fig:GVs}
\end{figure}

Electrical measurements were performed in a flow cryostat equipped
with a 14 T superconducting magnet and in a $^3$He cryostat. We
use standard lock-in techniques to measure $G-V$ curves, {\em
i.e.} curves of the differential conductance versus bias. In the
$G-V$ curves, at 4.2 K and below, a tunneling resonance is present
around 10 mV, superimposed on a dominant background. In
Fig.~\ref{fig:GVs} we show curves of a 400 $\mu$m device, for
temperatures between 0.5 and 12.5 K. The weak resonance at 4.2 K
becomes a clear peak at lower temperature, which is not yet
saturated at 0.5 K. At this temperature the FWHM of the peak is
about 1.5 mV. The overall behavior as sketched in
Fig.~\ref{fig:GVs} is present for all measured devices, which are
of different size and come from different fabrication runs. The
peak is absent for devices that have no $\delta-$layer but are
otherwise identical. This indicates that the peak originates from
the boron atoms of the $\delta-$layer.

\begin{figure}[t]
\includegraphics[width=7cm]{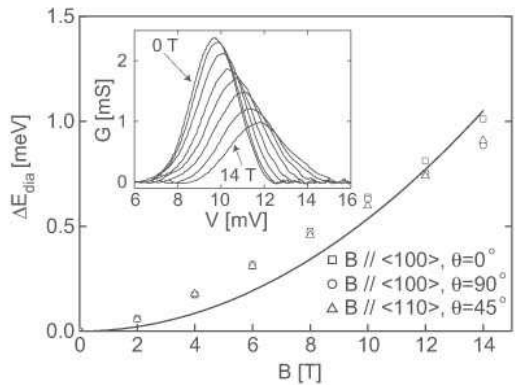}
\caption{Magnetic field induced shift of the discrete energy level
in the barrier of device 1/888/1, for three field orientations and
for three angles $\theta$ between field and current direction. The
parabola is a fit of the expression for the diamagnetic shift of
the B$^+$ level to the data points. The data points are derived
from curves as in the inset, which shows the field dependence of
the resonance of device 1/888/1 (the field step is 2 T).}
\label{fig:magnetic}
\end{figure}

In the inset of Fig.~\ref{fig:magnetic} we plot the response of
the spectra (background subtracted) to a magnetic field, for the
same device as in Fig.~\ref{fig:GVs}, for one bias
polarity~\cite{polarity} and for fields up to 14 T oriented
perpendicular to the layers. With increasing field the peak shifts
to higher bias and becomes broader and lower. In the main panel of
Fig.~\ref{fig:magnetic} data points of the level shift $\Delta
E=\frac{1}{2}e\Delta V_{\textrm{res}}$ deduced from the resonance
shift $\Delta V_{\textrm{res}}$ (see below for the relation
between level and resonance positions) follow a weak quadratic
function of the field, the shift at 14 T being about 1 mV. This
behavior does not depend on the orientation of the field with
respect to the crystallographic axes of the device.

The conductance peak is attributed to resonant tunneling through
the B$^+$ state of boron impurities in the $\delta-$layer, each of
which provides for a conductance channel. The B$^+$ state is an
acceptor counterpart of the more generally known D$^-$
state~\cite{DMLarsen}. It forms when a second hole is weakly bound
to a neutral acceptor, in our case neutral boron B$^0$. In zero
magnetic field the B$^+$ state is a singlet state (as the D$^-$
state is), which is analogous to the negative hydrogen
ion~\cite{DMLarsen} (H$^-$ ion). The separation of the B$^0$
ground-state level and the B$^+$ level results from the Coulomb
interaction between the holes. In our devices the B$^0$ ground
state is deep below the Fermi level, so that it is permanently
occupied. Therefore, higher levels of the B$^0$ single hole
spectrum are not available as stepping stones for holes tunneling
from emitter to collector. This means that the B$^+$ state is the
only candidate which can induce the resonance. This is unlike an
optical absorption experiment in which higher levels may always
come into play via transitions from the ground
state~\cite{Ramdas}.

\begin{figure}[t]
\includegraphics[width=6.5cm]{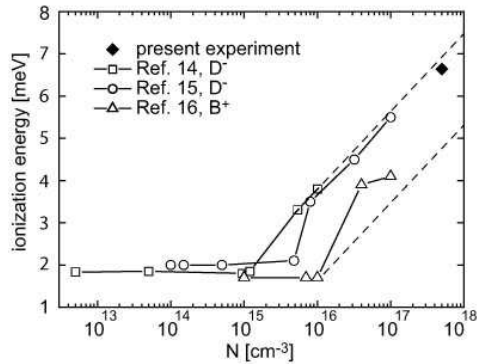}
\caption{Plot of experimental ionization energies for B$^+$ and
D$^-$ versus doping concentration (from literature). Above
$N\approx10^{15}$~cm$^{-3}$ the ionization energy increases. The
data point of the present work ($\blacklozenge$) nicely falls in
the extrapolated bandwidth which indicates the trend of the
literature data.} \label{fig:ionization}
\end{figure}

>From photoconductance spectroscopy on Si samples with low doping
level it is known~\cite{Gershenzon1} that in zero magnetic field
the binding energy $E^+(0)$ of the extra hole on an isolated B$^+$
ion, {\em i.e.} the amount of energy required to remove this hole
from the ion to the valence band edge, is about 2.0 meV. This is
close to 0.055Ry$^\ast=2.5$~meV predicted by the H$^-$ ion model,
which is successfully used for the D$^-$ state~\cite{DMLarsen}.
Here Ry$^\ast$ is the effective Rydberg for boron in silicon,
which equals the ground state energy $E_1=45.7$ meV~\cite{Ramdas}.
Since the measured resonance comes from the $\delta-$layer, equal
parts of the bias drop across the barriers at either side of the
Coulombic potential well associated with a boron impurity. Thus,
the resonance voltage is $V_{\textrm{res}}=2[\phi_B-E^+(0)]/e$
[see Fig.~\ref{fig:SIMS}(b)]. We determine $\phi_B$ from the
temperature dependence of the zero-bias resistance $R_0$. When
plotted versus $1/T$, the logarithm $\ln{(T^2R_0)}$ clearly shows
activated behaviour in the range 15-20 K. Interpreting this as
Richardson-Dushman thermionic emission~\cite{Kao} of holes over
the barrier, we find $\phi_B=11.7$~meV from a fit to the data.
This is not too far from $\phi_B=\Delta E_v-E_F=8$~meV, which we
obtain from the estimate~\cite{Jain} for $\Delta E_v(N_B)$ and
from photoluminescence measurements~\cite{Dumke} of $E_F(N_B)$,
each subject to uncertainty. The values $\phi_B=11.7$~meV and
$E^+(0)=2.0$~meV predict $V_{\textrm{res}}=19.4$~mV. This deviates
strongly from the measured position and thus seems to exclude the
B$^+$ state. However, the concentration
$N_B=5\times10^{17}$~cm$^{-3}$ of the $\delta-$layer is high
enough for the tail of the wave function of the second hole to be
appreciable at the nearest B$^0$ atoms~\cite{Norton}, so that the
B$^+$ ions are not isolated. The additional Coulomb attraction of
these B$^0$ atoms and spreading of the electron charge among them
(reducing the hole-hole repulsion at the B$^+$ state) cause a
stronger binding. This effect increases with increasing
concentration. The measured $V_{\textrm{res}}\approx 10$~mV gives
$E^+(0)\approx 6.7$~meV. As demonstrated in
Fig.~\ref{fig:ionization}, this value nicely falls in the
bandwidth obtained by extrapolating the scarce experimental
data~\cite{Norton,Taniguchi,Burger} on the concentration
dependence of the ionization energy for B$^+$ and D$^-$.

In the lower part of the temperature range of Fig. 2 the resonance
is much wider than the theoretical width $3.5kT$ of a
Fermi-smeared sharp resonance, implying a rather large zero
temperature width ($\approx1.5$~meV). Mechanisms contributing to
this are life-time broadening, disorder broadening and broadening
due to the finite width of the $\delta-$layer. Life-time
broadening determines the intrinsic width of a resonance coming
from a single impurity. For the B$^+$ level life-time broadening
is appreciable, since it is so close to the ionization level. The
effect is enhanced by the large bias field at resonance (2.5
kV/cm), which weakens the collector barrier and thus shortens the
life time. We will come back to this in discussing broadening of
the resonance with increasing magnetic field (see below). Disorder
broadening arises from the different local surrounding of B$^0$
centers by adjacent B$^0$ centers, which causes a distribution of
levels. The finite width of the $\delta-$layer implies a range of
values of $V_{\textrm{res}}$, since the position of the atom
defines the barrier thicknesses and thus the distribution of the
bias over them. Without correcting for a decay of the resonance
for off-center positions~\cite{Kalmeyer}, a range for
$V_{\textrm{res}}$ of 3.5 mV is derived from only the finite
$\delta-$layer width. We take this as a sign that the
$\delta-$layer width is an important source of broadening.

The resonance shift to higher bias (Fig.~\ref{fig:magnetic})
reflects the diamagnetic shift $\Delta E_{\textrm{dia}}(B)$ of the
B$^+$ level in a magnetic field. This shift, termed diamagnetic
because of the related negative susceptibility, is towards the
valence band edge, in agreement with the observed peak shift to
higher biases. In magneto-optical spectroscopy the diamagnetic
shift is not obtained directly, since it is absorbed in the
field-dependent binding energy
$E^+(B)=E^+(0)+\frac{1}{2}\hbar\omega-\Delta E_{\textrm{dia}}(B)$.
Here $\frac{1}{2}\hbar\omega$ is the energy of the first Landau
level, which is the valence-band edge in field. In our experiment,
however, since the emitter doping level $N_B=10^{19}$~cm$^{-3}$ is
high enough to block Landau level formation in this layer, the
emitter Fermi level is a field-independent reference energy
enabling direct measurement of $\Delta E_{\textrm{dia}}(B)$. For
the weak fields used here, first order perturbation theory
estimates the diamagnetic shift as~\cite{Landau} $\Delta
E_{\textrm{dia}}(B)=\frac{e^2B^2}{12m}~\sum_i \overline{r_i^2}$.
Here $m$ is the effective mass of the holes of the B$^+$ ion and
$\overline{r_i^2}$ is the mean square distance of the $i$-th hole
to the B$^-$ core ($i=1,2$). Taking~\cite{holemass}
$m_{lh}=0.15m_0$ and $m_{hh}=0.54m_0$ for the light and heavy hole
hole masses, respectively, and $\overline{r_1^2}=a_0^2$ for the
first hole ($a_0$=3.9 nm is the Bohr radius of the $B^0$ atom), we
fit the expression for $\Delta E_{\textrm{dia}}(B)$ to the
complete set of data points of Fig.~\ref{fig:magnetic}. This
yields the fit shown in the figure and corresponding values
$\overline{r_2^2}=(1.6 a_0)^2$ and $\overline{r_2^2}=(3.4 a_0)^2$.
The range defined by these values is consistent with $r_2=2.4 a_0$
cited for the B$^+$ state~\cite{Gershenzon2}. We attribute the
deviation of the fitted curve from the experimental trend to the
use of an atomic physics model for $\Delta E_{\textrm{dia}}$ in a
solid state system.

Broadening of the resonance with increasing magnetic field may be
unresolved  splitting and/or life-time broadening induced by the
Stark effect. Since the resonance voltage increases with
increasing magnetic field, the electic field at resonance, and
thus the Stark broadening, increase as well. Data for Stark
broadening of the $B^+$ level are not available. Therefore, we
take as a measure the broadening of the far infrared absorption
line due to transition from the ground state of $B^0$ atom in
silicon to the first excited state, which was measured up to 1.0
kV/cm~\cite{White}. Extrapolation of the data of Ref.~[21] to the
fields of our experiments yields an increase of the halfwidth
between 0 T and 14 T of 0.1 meV, to be compared with our measured
increase of 0.5 meV. Stark broadening apparently plays a role.

Finally, we discuss the background contribution to the conductance
(see Fig. 2), which shows a weak temperature dependence below 4.2
K. The parabolic shape of the background suggests direct tunneling
as transport mechanism. However, the conductance at $V=0$ is
several orders of magnitude higher than expected for direct
tunneling~\cite{Meservey}, so that it is excluded. Hence, the
background conductance is due to hopping resulting from the
background doping in the barrier. This hopping is thermally
activated close to $V=0$ and field activated at higher biases
($|V|\ge2~$mV). For biases exceeding the barrier height
($|V|\ge11.7$~mV), the barrier becomes increasingly weak for
hopping and finally for direct tunneling, giving a further
conductance increase. Above 4.2~K the conductance in the range
($|V|\le2~$mV) undergoes a transition to coexistence of thermal
hopping and thermal activation over the barrier and finally to
dominance of activation over the barrier.

In conclusion, we have studied resonant tunneling through a Si
barrier $\delta-$doped with boron impurities. The conductance
resonance observed is due to tunneling through the B$^+$ state of
the impurities. The structure of our device enables direct
observation of the diamagnetic shift of the B$^+$ state. The
measured magnitude of the shift agrees well with the theoretical
description, yielding a proper value of the orbit size for the
second hole of the B$^+$ state.  The binding energy of the B$^+$
state turns out to be enhanced as a result of overlap of the wave
function of the second hole of the $B^+$ state with the nearest
boron impurities. Our next step will be miniaturizing the devices
to the level of one dopant atom (diameter$\approx$50 nm), enabling
studies of the effect of wave-function manipulation on transport
through the atom.

We acknowledge valuable discussions with G.E.W. Bauer, T.O.
Klaassen and J.R. Tucker. M. van Putten is acknowledged for his
contributions in the initial phase of the work. This work is part
of the research program of the Stichting voor Fundamenteel
Onderzoek der Materie, which is financially supported by the
Nederlandse Organisatie voor Wetenschappelijk Onderzoek. One of
us, S.R., wishes to acknowledge the Royal Netherlands Academy of
Arts and Sciences for financial support.


\begin{thebibliography}{99}
\bibitem{polarity} The conductance curves are very symmetric, so
that it suffices to measure for only one bias polarity.
\end{thebibliography}
\end{document}